\documentclass[twocolumn,prl,aps,epsf,showpacs]{revtex4}
\usepackage{epsfig}

\begin{document}

\title{Search for Free Decay of Negative Pions in Water and Light Materials}

\author{T. Numao}
\affiliation{TRIUMF, 4004 Wesbrook Mall, Vancouver, B.C., Canada V6T 2A3}

\author{T.C. Awes}
\affiliation{Oak Ridge National Laboratory,
High Energy Physics, Oak Ridge, TN 37831-6372}

\author{S. Berridge}
\affiliation{University of Tennessee, Knoxville,
TN 37996-1200}

\author{W. Bugg}
\affiliation{University of Tennessee, Knoxville,
TN 37996-1200}

\author{V. Cianciolo}
\affiliation{Oak Ridge National Laboratory,
High Energy Physics, Oak Ridge, TN 37831-6372}

\author{Yu.I. Davydov}
\affiliation{TRIUMF, 4004 Wesbrook Mall, Vancouver, B.C., Canada V6T 2A3}

\author{Yu. Efremenko}
\affiliation{University of Tennessee, Knoxville,
TN 37996-1200}

\author{R. Gearhart}
\affiliation{University of Tennessee, Knoxville,
TN 37996-1200}

\author{S. Ovchinnikov}
\affiliation{University of Tennessee, Knoxville,
TN 37996-1200}

\author{J-M. Poutissou}
\affiliation{TRIUMF, 4004 Wesbrook Mall, Vancouver, B.C., Canada V6T 2A3}

\date{\today}

\begin{abstract}

We report on a search for
the free decay component of $\pi^-$ stopped in water
and light materials.
A non-zero value of this would be
an indication of anomalous $\overline{\nu}_e$ contamination
to the $\nu_e$ and $\overline{\nu}_{\mu}$ production
at stopped-pion neutrino facilities.
No free decay component of $\pi^-$ was observed in water, Beryllium,
and Aluminum, for which upper limits were  established at 
$8.2 \times 10^{-4}$, $3.2 \times 10^{-3}$, and $7.7 \times 10^{-3}$,
respectively.

\end{abstract}

\pacs{14.60.Pq, 29.25.-t, 36.10.-k, 25.80.Hp}

\maketitle

\section{I. Introduction}

Flavor suppression in the incident neutrino beam is one of the key elements
for successful
neutrino appearance experiments.
In the case of beam-dump experiments using muon decays from stopped pions, 
the contribution to neutrino production from
negative pions is expected to be significantly suppressed, because
negative pions are believed to undergo nuclear capture promptly after stopping in
material, while positive pions freely decay.
The ratio of $\pi^-$/$\pi^+$  production cross sections for a proton beam
in the 0.5--1 GeV energy region is about 1/5 for light
nuclei, which further enhances
this asymmetry.
The total neutrino content after the pion decay
$\pi^+ \rightarrow \mu^+ \nu_{\mu}$ followed by the muon decay
$\mu^+ \rightarrow e^+ \nu_e \overline{\nu}_{\mu}$
($\pi^+ \rightarrow \mu^+ \rightarrow e^+$ decay) does not include
$\overline{\nu}_e$'s in the final state.
The contribution from decay-in-flight is a few per cent for a pion. 
This provides a good opportunity
for $\overline{\nu}_e$ appearance experiments.\cite{lsnd,karmen,new}
However, a finite $\pi^-$ lifetime in the stopping material
may result in additional $\overline{\nu}_e$ contamination.

Normally, after stopping and being trapped in an atomic orbit, a negative pion orbiting 
the nucleus promptly de-excites by the Auger process,
radiative cascade decays, or Stark mixing to an S state, 
where the significant overlap of the wave function with the nucleus
enables the $\pi^-$ to be captured by the nucleus.
If the $\pi^-$ is not captured by the nucleus it may decay freely in the atomic state.
The free decay fraction of $\pi^-$'s was measured to be
$8.7 \times 10^{-5}$ \cite{hydrogen} in Hydrogen and is expected to decrease with the atomic number.
However, there is an exception in the $\pi^-$He atom system\cite{helium,nakamura}, where
the $\pi^-$ is trapped in a meta-stable state that has little wave-function
overlap with the nucleus and the
Auger processes are suppressed\cite{condo}. 
It is therefore important to confirm the absence of other mechanisms that may cause
free decays of $\pi^-$'s.

Limits for the fraction of the free-decay component 
exist for H$_2$ and He\cite{hydrogen,helium,nakamura} 
(data in Ne is only applicable to a meta-stable state
with a lifetime longer than $\sim 1$ ns).
This paper reports the results of a search for a free decay component
of $\pi^-$ in water and light materials from  analyses of electron yields and
proton spectra.
\\

\section{II. Principle}

Figure 1 illustrates
the time diagrams involved in pionic atom formation (left)
up to the disappearance of the pion
by (A) prompt nuclear capture,
 (b and B) free decay with a partial lifetime  
$\tau_{\pi}$, and
(C) delayed pion capture with a partial lifetime of $\tau_{capt}$.
The processes (a) and (c) feed into  ``$\pi^-$ Atom'' (right).
Since the diagrams for delayed pionic atom formation (left)
and delayed nuclear capture (right) are identical, 
and since the two processes cannot be discriminated or 
separated in this analysis,
the two processes are considered as a single process 
with prompt capture notation,
$i.e.$ $R$ and $\tau_{capt}$ are used in this analysis.  
The common observed lifetime, 
$\tau_{obs} = 1/(\frac{1}{\tau_{\pi}} + \frac{1}{\tau_{capt}})$,
in the diagrams already assumes only delayed capture process.

\begin{figure}[htb]
\begin{minipage}[t]{86mm}
\begin{center}
\epsfig{file=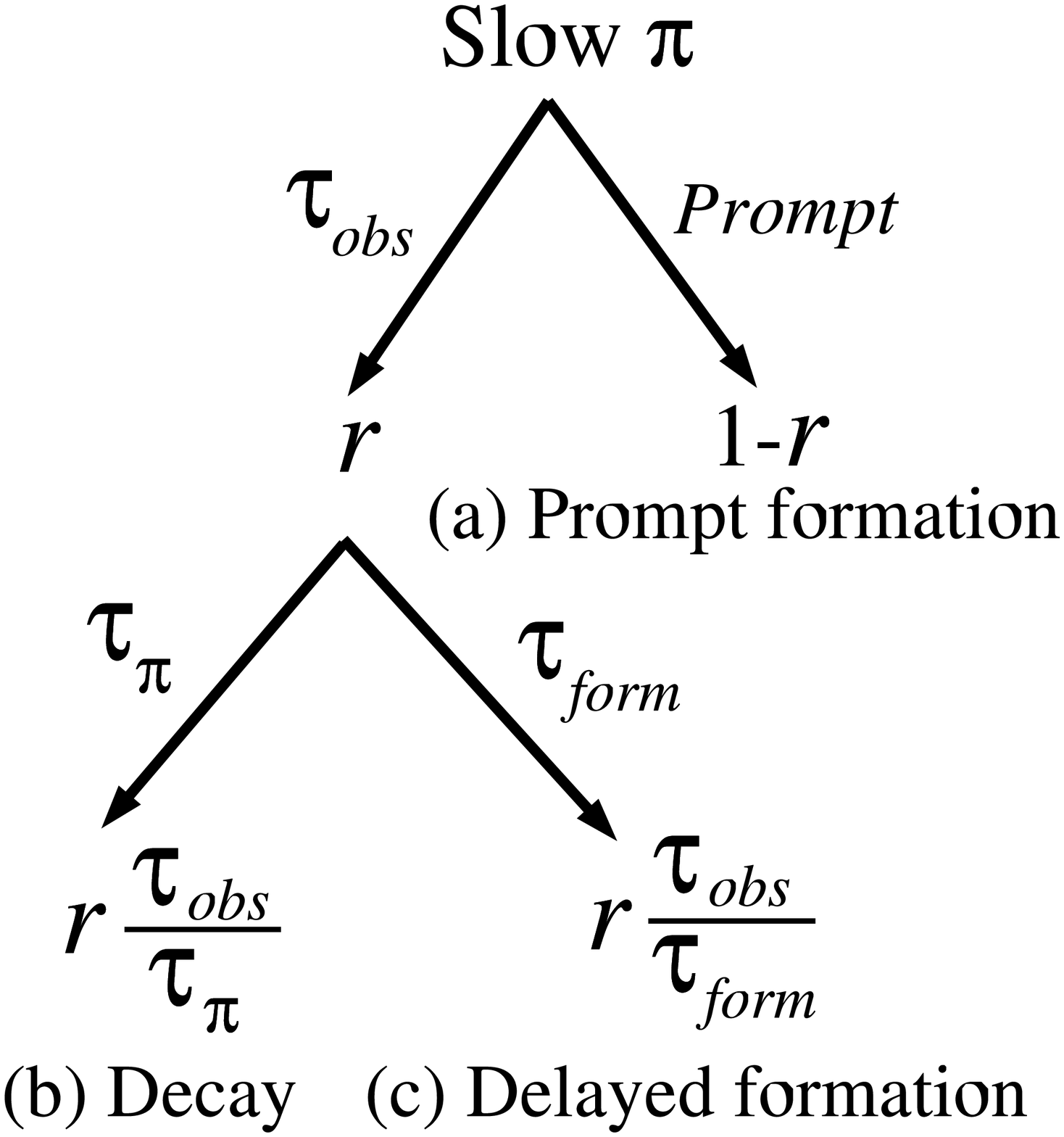,width=4.2cm,clip=}
\epsfig{file=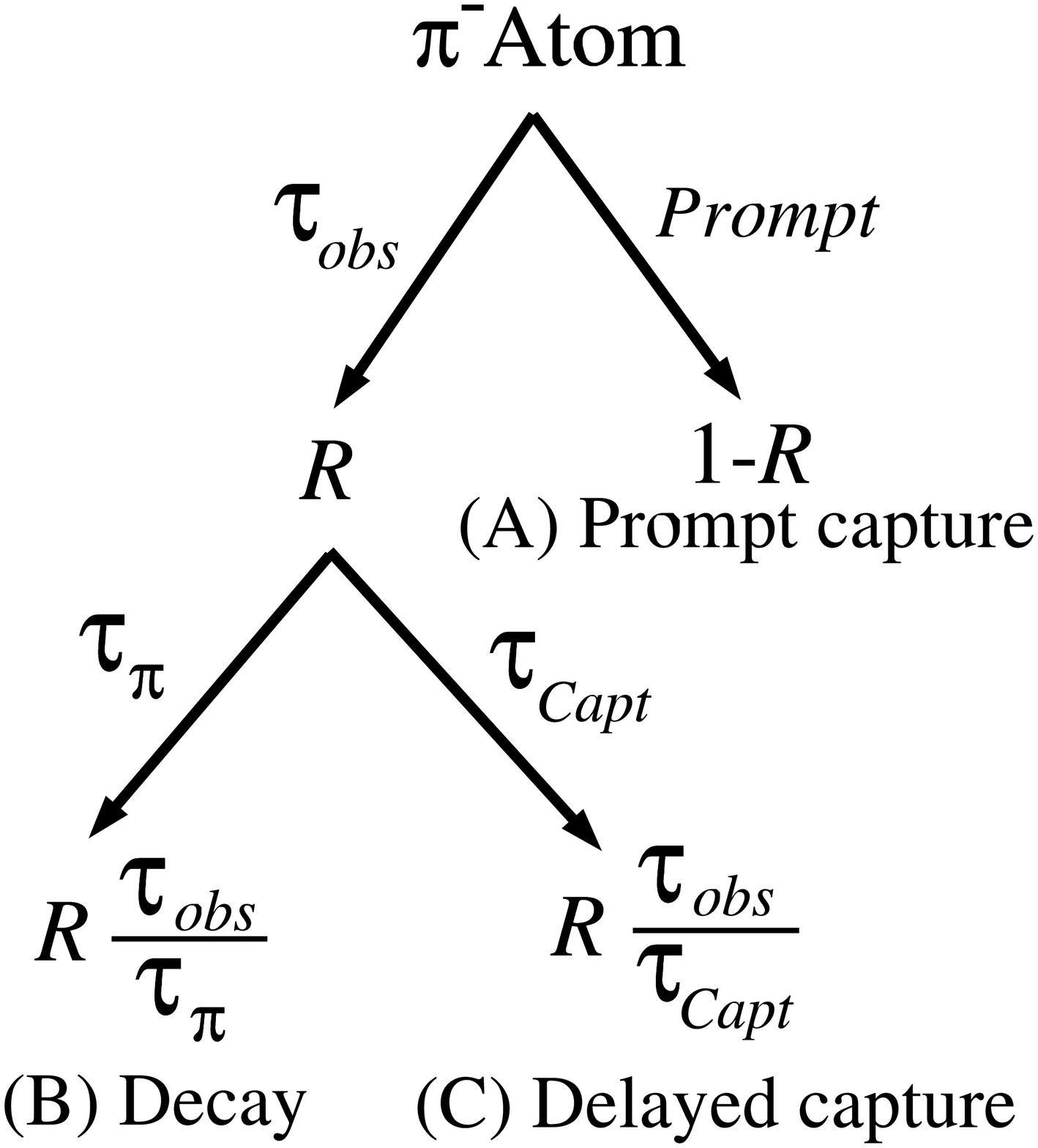,width=4.2cm,clip=}
\caption{Left: Pionic atom formation processes, 
and Right: Atomic de-excitation and capture processes of the pionic atom.
$r$ and $R$ are the fractions of delayed pionic-atom
 formation and delayed capture,
respectively. $\tau_{form}$ and $\tau_{capt}$ indicate
 the partial lifetimes for
delayed pionic-atom formation and delayed capture, respectively.
}
%\label{fig:largenenough}
\end{center}
\end{minipage}
\end{figure}

A direct method to search for free $\pi^-$ decay is to detect decay products of pions (B). 
Detection of muons from the decay 
$\pi^- \rightarrow \mu^- \overline{\nu}$ in water is not practical
because of the short muon range in water ($\sim 1.4$~mm).
However, electrons from the $\pi^- \rightarrow \mu^- \rightarrow e^-$ decay chain
($E_e = 0-53$~MeV) can be detected.
In this method,
the background from decay-in-flight (DIF) of a pion to a muon,
which stops in materials around the target and later decays to an electron,
must be subtracted using Monte Carlo
(MC) calculations.
The advantages of
the search for electrons from the decay
$\pi^- \rightarrow \mu^- \rightarrow e^-$ are that
the measurement is the sum of all free decay components with different lifetimes
and that the sensitivity 
 does not strongly depend on the lifetime $\tau_{obs}$;
it varies only by $\sim$10 \% 
for the lifetime between 0 ns and the pion lifetime 
$\tau_{\pi} = 26.033$~ns\cite{pilife}.

Alternately, free $\pi^-$ decay may be detected by the presence of  
electrons from the decay $\pi^- \rightarrow e^- \nu$.
However, because of its small branching ratio, 
$1.231 \times 10^{-4}$ \cite{pienu},
it is very
difficult to differentiate it from the background arising
from prompt nuclear capture $\gamma$-rays that convert in the target to an
electron-positron pair. The search for $\pi^- \rightarrow e^- \overline{\nu}$
is effective for a component with a relatively long lifetime.

The third method (indirect) 
is to search for delayed components (C) in the
time spectrum of pion capture products, such as protons, $\gamma$-rays,
and $\pi^0$'s.
The ratio of delayed $\pi^-$ capture $f_{capt}$ is expressed as

\begin{equation}
f_{capt}=\frac{\mbox{(C)}}{\mbox{(A)+(C)}}
=\frac{R \frac{\tau_{obs}}{\tau_{capt}}}{1-R+R \frac{\tau_{obs}}{\tau_{capt}}}.
\end{equation}

\noindent
%Using directly observable parameters,
By eliminating $R$ and $\tau_{capt}$, which are not directly observable,
the free decay fraction (B)  can be deduced,

\begin{equation}
\mbox{(B)} = R \cdot \frac{\tau_{obs}}{\tau_{\pi}}
= \frac{f_{capt} \cdot \tau_{obs}}{\tau_{\pi}-\tau_{obs} +
f_{capt}\cdot \tau_{obs}}.
\end{equation}

\noindent
It should be noted that when $\tau_{obs}$ approaches $\tau_{\pi}$
the equation approaches to unity, indicating that
the capture experiment is insensitive to the free decay component
as $\tau_{capt}$ approaches infinity.
In order to be sensitive to
a fast decay component in the presence of the
dominant prompt capture (the process (A)), the experiment is  required  
to have good time resolution.
In the search for delayed components in water,
this can be achieved by measuring protons that have high energy losses,
 therefore, with small time walk and spread.
Proton measurements  are not sensitive to $\pi^-$ capture by Hydrogen
in water,
since only $\gamma$-rays, $\pi^0$'s, and neutrons are produced.
The majority ($>99$ \%) of $\pi^-$'s in water\cite{water,water2} 
are captured by Oxygen atoms.

\begin{figure}[htb]
\begin{minipage}[t]{86mm}
\begin{center}
%\framebox[79mm]{\rule[-26mm]{0mm}{52mm}}
\epsfig{file=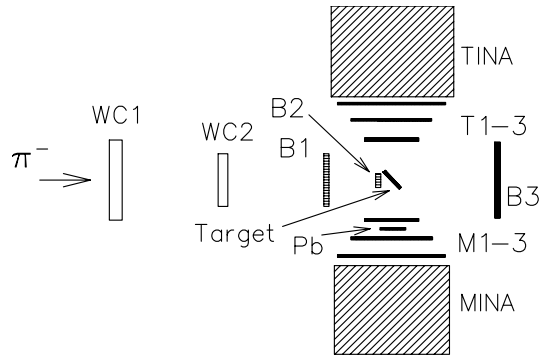,angle=90,width=7cm,clip=}
\caption{Experimental setup.
TINA and MINA are NaI(Tl) detectors described in the text.}
%\label{fig:largenenough}
\end{center}
\end{minipage}
\end{figure}

\section{III. Experiment}

The experiment was performed at the TRIUMF M9A channel.
Typical ratios of incoming beam particles 
at the experimental target were e:$\mu$:$\pi$=5:1:4 for a positive-particle beam
and  e:$\mu$:$\pi$=88:3:9 for a negative-particle beam
at $P_{\pi} = 70$~MeV/c with a momentum bite $\Delta P/P = \pm 4$~\%.
Figure 2 shows the detector set-up.
The position and the direction of incoming beam particles were measured 
by two sets of X-Y read-out wire chambers (WC1 and WC2) located  95 cm and 50 cm
upstream of the target. 
Two plastic scintillators  B1 (76-mm high, 76-mm wide,
3.2-mm thick) and B2 (35-mm high, 25-mm wide, 5-mm thick) were placed
8 cm and 3 cm upstream of the target center, respectively.
 In order to collimate the beam
(to detect scattered beam by $e.g.$ the WC frames), a $305 \times 305$mm$^2$ plastic
counter with a 76-mm diameter hole (not shown in the figure)
was placed 5 cm upstream of the
B1 counter. Purified water for the target was held in a $60 \times 60 \times 8$ mm$^3$ container
made of 0.2-mm thick aluminum frame 
and 50-$\mu$m thick beam windows (both ends).  
The target was
tilted to 45 degrees with respect to the beam (some runs were taken at --45 degrees). 
A $152 \times$152-mm$^2$ veto counter B3
for the detection of passing-through beam particles
was placed 40 cm downstream of the target.
More than 90 \% of pions stopped in the target
at a rate of 1--10 kHz for the $\pi^-$ beam and 30--100 kHz for $\pi^+$.

The decay products were observed by two telescope arms at 5 cm from the beam axis
at $\pm 90$ degrees.
Each telescope consisted of three 3--6 mm thick plastic scintillators
(T1--T3 or M1--M3), and a 46-cm diameter,
51-cm long NaI(Tl) crystal (TINA) or a 36-cm diameter, 36-cm long NaI(Tl) (MINA).
In each telescope, the farthest scintillator from the target was the largest 
one that covered the front face
of the NaI for pile-up detection.
The solid angle of each telescope was defined by the 102 $\times$ 102-mm$^2$ T1
or the 76 $\times$ 76-mm$^2$ M1 counter.
For $\pi^0$ detection, a 5-mm thick $51 \times 51$-mm$^2$ lead sheet was placed
as a converter
between the first and the second plastic scintillators in the MINA telescope.
\begin{figure}[htb]
\begin{minipage}[t]{86mm}
\begin{center}
%\framebox[79mm]{\rule[-26mm]{0mm}{52mm}}
\epsfig{file=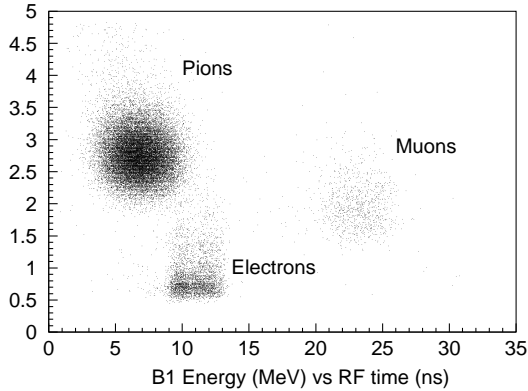,width=7cm,clip=}
\caption{Energy in B1 vs beam burst (RF) time with respect to B2 time.
The time walk effect in the B2 counter has been corrected.}
%\label{fig:largenenough}
\end{center}
\end{minipage}
\end{figure}
\begin{figure}[htb]
\begin{minipage}[t]{80mm}
\begin{center}
%\framebox[79mm]{\rule[-26mm]{0mm}{52mm}}
\epsfig{file=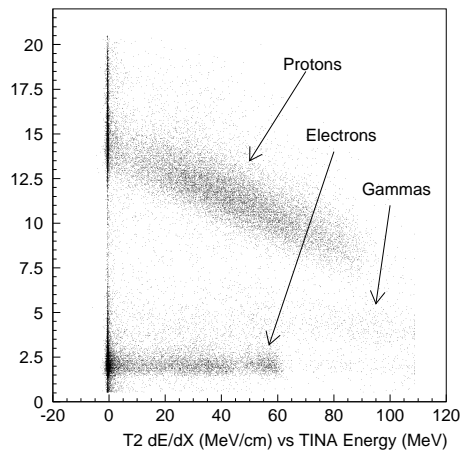,width=6cm,clip=}
\caption{Energy loss in T2 (MeV/cm) vs energy in TINA (MeV).
Electrons below 45 MeV are from $\pi \rightarrow \mu \rightarrow e$ decays and
around 60 MeV are from electron scattering.}
%\label{fig:largenenough}
\end{center}
\end{minipage}
\end{figure}

The trigger required a pion stopping in the target (B1$\cdot$B2$\cdot \overline{\mbox{B3}}$) 
with the presence of an outgoing particle into the T2 or M2 counter
in the time window from --50 ns to 150 ns
with respect to the prompt time---this condition allowed
to accept events containing electrons, $\gamma$-rays, $\pi^0$'s (two $\gamma$-rays),
 protons, and deuterons.  
Typical trigger rates were 70 Hz for $\pi^-$  runs  and 150 Hz for $\pi^+$
with a deadtime of 10--30 \%.
The timing and pulse heights of all the scintillation counters,
two coordinates from each wire chamber,
 and the relative phase of the 23.06-MHz cyclotron RF 
(incident  proton beam bunches)
to the B2 counter time for the time-of-flight (TOF) measurement of a beam particle 
 were recorded.
In order to record pile-up information, additional wide-gate (250 ns) ADC's were
employed using signals split from B1, B3, T3, and M3 counters for detection of extra charge
outside the standard gate (50 ns). Pile-up in the NaI counters was measured by using two
additional ADC's with an early gate that closed just before the main pulse, 
and a late gate 
(by 200 ns) to compare to the normal gate ADC.

Special runs were taken for calibration and normalization;
runs with a trigger using a pulser driven by a pre-scaled cyclotron RF signal,
runs requiring only the presence of incoming particles (B1$\cdot$B2),
and runs with positive beam particles (for normalization of the electron
acceptance).
For  $\pi \rightarrow \mu \rightarrow e$ measurement, a 30-mm diameter,
12-mm thick cylindrical water target
 was used without tilting, together with a 25-mm diameter, 5-mm thick B2 counter
that faced the target with a variable gap between $z = 0-3.5$~cm.
Also, other candidate materials for the beam moderators at a neutrino factory,
Beryllium ($50 \times 50 \times 8$ mm$^3$) and Aluminum ($75 \times 75 \times 6$ mm$^3$),
were used as targets during several runs.  A vinegar target (5 \% acetic acid by volume)
 was used to simulate
possible effects of excessive H$^+$ ions
in water generated by the high-intensity proton beam of the stopped-beam facility.
Measurement was made  at beam momenta of  64, 67, 70, 73, 76 and 80 MeV/c to confirm
the stopping position of pions and to test systematic effects.
\\

\section{IV. Analysis}

Beam pions were identified by the energy loss in the first beam counter
B1, and the TOF with respect to the proton beam bunch.
Figure 3 shows a scatter plot of B1 counter energy vs RF time with respect to the
B2 time after the time walk correction.
At the experimental target position (about 10 m from the pion production
target), the TOF of electrons coming from the next beam bucket was
separated from that of 70-MeV/c pions,
and the probability of misidentifying a beam electron as a pion was
estimated to be $\sim 2 \times 10^{-7}$.  The TOF peak of
muons was clearly separated.

The beam counters were calibrated using estimated energy losses of
$e$'s, $\mu$'s and $\pi$'s
that passed through the counters.
The telescope counters were calibrated with positrons from
$\mu^+ \rightarrow e^+ \nu \overline{\nu}$ decays, 
$\pi^+ \rightarrow e^+ \nu$ decays and
scattered beam particles.
For TINA and MINA, energy losses in the telescope counters and the 1-MeV shift 
due to positron annihilation
were taken into account.
Time calibration was based on the repeated peaks with the RF cycle 
in the time spectra.
Figure 4 shows a scatter plot of energy in the
T2 counter vs energy in TINA,
in which the diagonal band is from protons produced by pion capture,
the flat distribution around
the energy loss of 2 MeV/cm
is from electrons (up to 45 MeV from 
$\mu \rightarrow e \nu \overline{\nu}$ decays
and around 60 MeV from scattered beam electrons), and the distribution
around 100 MeV with twice the energy loss of electrons in the T2 counter is from 
electron-positron pairs from $\gamma$-rays
converted in the T1 counter or the target.

\subsection{A. Analysis of $\pi \rightarrow \mu \rightarrow e$ decay}

In the analysis of $\pi^- \rightarrow \mu^- \rightarrow e^-$ decay events,
data taken with the circular water target were used.
Pulse heights in the 
three telescope counters T1--T3 were required to be less than $\sim$7 MeV/cm.
The effect of the loss by this cut ($\sim$3 \%)
 cancelled
when the yield was normalized to that 
of $\pi^+ \rightarrow \mu^+ \rightarrow e^+$.
No proton contamination below $E$=80 MeV  was observed.
Some events from two ``electrons''
were accepted with
the $dE/dx$ cut in the T-counters, but
they appeared together with scattered beam electrons
at the prompt time and at every 43 ns with 
respect to the prompt time.
Only delayed events 
 outside those affected regions 
($t_e$ = 48--81 ns with respect to the prompt time $t_0$)\cite{early}
were selected for the analysis.
\begin{figure}[htb]
\begin{minipage}[t]{86mm}
\begin{center}
%\framebox[79mm]{\rule[-26mm]{0mm}{52mm}}
\epsfig{file=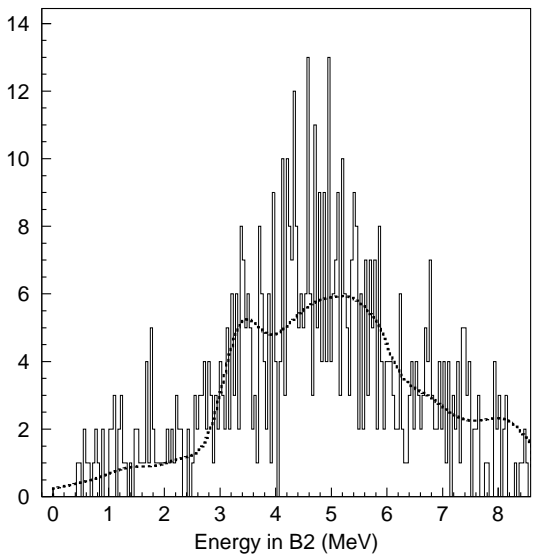,width=7cm,clip=}
\caption{Energy spectrum measured in the B2 counter.
The dotted line shows the result of MC calculation normalized to the
total count in the spectrum.}
%\label{fig:largenenough}
\end{center}
\end{minipage}
\end{figure}

The histogram of 
Figure 5 shows the energy spectrum measured in the B2 counter
for the events that passed the above cuts.
The position of the B2 counter was in contact with
the target ($z$=0 cm).
DIF contributions before and after passing the B2 counter
are the major remaining components.
Events from decays after passing the B2 counter form
a peak around 5 MeV that corresponds to energy loss of pions
passing through the B2 counter prior to stopping in the target.
Events from the decays
before passing the B2 counter form
a broad distribution with a shoulder 
around 3.5 MeV that corresponds to the energy loss of 
the highest energy muons ($\sim$32 MeV for forward pion decays) and
the lower energy 
events are from the backward-decay muons that stopped in the B2 counter.

%The contribution of DIF events was simulated by MC.
The contribution of DIF events was simulated by MC based on GEANT3 \cite{geant3}
with a modification for the $\pi^-$ treatment.
Pions were generated at the exit of the beam pipe (110 cm upstream of the target)
with a profile of $10 \times
10$ cm$^2$ projecting to the target ($2 \times 2$ cm$^2$) according to
a flat momentum distribution centered at 70 ($\pm$ 4 \%) MeV/c that reproduced
the measured energy spectrum of pions taken with a beam trigger.
When a $\pi^-$ stopped in material, it was promptly
  captured (capture products were not generated in the $\pi \rightarrow \mu \rightarrow e$
  simulation).
  Muon capture was not included in the MC, but the effect
  was included in the lifetime correction by using the free muon lifetime
  instead of the lifetime of the muonic atom \cite{muatom}.
  The effect of the binding energy (0.2 MeV for the muonic Oxygen atom) on the
  electron spectrum was ignored.
The  MC events were analyzed with the same selection criteria as the 
experimental data and the
energy spectrum in the B2 counter was obtained as shown by the dotted line in Fig. 5
(normalized to the total count).
The shape of the distribution is well reproduced by MC calculations
of the DIF events described above.
The ratio of the two components in the MC spectrum, the peak (DIF after B2)
and the broad bump (DIF before B2), strongly depended on the gap $z$, especially
near the target ($z$=0 cm).
Although the peak is, in principle,  more sensitive to the free decay component,
the sum was used in the
analysis because of
this strong dependence near $z$=0 cm for reliable subtraction of the background
estimated by the MC calculations.
The contamination from the $\mu$'s in the beam was negligible. 

In order to obtain the free decay component fraction,
the yield was normalized to that of the $\pi^+ \rightarrow \mu^+ \rightarrow e^+$
decays in the same time window with the corrections for the lifetime and the stopping
fractions  obtained from the beam-trigger data 
($f_{\pi^+}:f_{\pi^-} = 0.81:0.69$) \cite{stop}.
The second column in
Table I shows the gap dependence of the free decay fractions,
which are dominated by the DIF contribution.
For the normalization of the MC calculations,  
$\pi^+$ data were also generated and analyzed
in the same way.
The last column in
Table I summarizes the corresponding fractions from the MC calculations.

\begin{table}
\begin{center}
\begin{tabular}{|l|l|l|}
\hline
Gap (cm) & \multicolumn{2}{c|}{Free decay fractions} \\
\cline{2-3}
 &Measured (\%) & MC (\%) \\
\hline
0 & 1.12 $\pm$ 0.04 & 1.09 $\pm$ 0.03 \\
  & 1.09 $\pm$ 0.04 & \\
%0.5 & & 1.12 $\pm$ 0.04\\
1.4 & 1.26 $\pm$ 0.05 & 1.17 $\pm$ 0.04 \\
%2.5 & & 1.06 $\pm$ 0.04\\
3.5 & 1.15 $\pm$ 0.04 & 1.07 $\pm$ 0.03 \\
\hline
\end{tabular}
\end{center}
\caption{Summary of  $\pi \rightarrow \mu \rightarrow e$ analysis
for the water target. To obtain the free decay fractions 
for the experimental data and the MC calculations, the yields
have been normalized to those from $\pi^+$, 
and the time-window correction has been applied.}
\end{table}

The two measurements at $z=0$~cm  represent data with a possible
  beam position shift due to the magnet setup adjustment to correct for the
  hysteresis effect.
The average of two measurements at $z$=0~cm was used as the final result.
An upper
limit based on Bayesian method assuming a Gaussian probability distribution was
obtained by subtracting the MC result and adding the associated errors
in quadrature, and then renormalizing the area above zero.
The obtained 90 \% Confidence Level (C.L.) upper limit of the $\pi \rightarrow \mu \rightarrow e$
amplitude corresponds to
free-decay fraction of  $7.5 \times 10^{-4}$ for a short lifetime
and  $8.2 \times 10^{-4}$ for a lifetime of the free-decay 
component near $\tau_{obs} = \tau_{\pi}$.
\\

\subsection{B. Analysis of $\pi \rightarrow e \nu$ decay}

The signal region was defined by a box cut of
$t_e$=5--100 ns and $E_e$=60--72 MeV in TINA.
Event selection criteria for the $\pi \rightarrow e \nu$ decay analysis
were tighter than those used in the $\pi \rightarrow \mu \rightarrow e$
analysis because of its small branching ratio and sensitivity to background.
A typical type of background occurred when a pion stopped in the target and
an additional beam electron was scattered into TINA.
Protection against this came from efficient beam pile-up detection and
the high TINA energy threshold---
 the presence of extra beam particles
during the search time window
was detected by tighter energy loss cuts in the B1 and B2 counters as well as
by comparing the pulse charges observed in the narrow-gate and wide-gate ADC's
for the B1 counter.
Using the observed pile-up data and beam electron energy spectra,
the level of this background in terms of
free decay fraction was estimated to be $4 \times 10^{-3}$.
Pile-up in TINA that shifts the energy of  $\pi \rightarrow \mu \rightarrow e$
events into the signal region was another concern. 
Early pile-up in TINA was detected by the ADC with an early gate that closed at the
beginning of the normal signal.
Other pile-up was detected by a set of two ADC's with different gate widths
for TINA and T3.
This background was $<$ 1 \%
of $\pi^+ \rightarrow e^+ \nu$ decays and significantly 
suppressed for $\pi^-$ by the lack of 
free pion decays.

Figure 6 shows energy spectra in TINA for $\pi^+$ (top) and $\pi^-$ (bottom) runs 
with the water target.
Some  events were seen in the time-energy box with time distribution consistent 
with the electron scattering background,
though they were included in the free-decay candidates.
90 \% C.L. upper limits were obtained assuming Poisson statistics
with no background.
The limits were normalized to the corresponding $\pi^+$ data with the time
window corrections.
The results are summarized in Table II.
The 90 \% C.L. limits shown in the Table correspond 
to the case when the free decay component
has the pion lifetime.
These limits are complementary to the capture measurements. 
\\
\begin{figure}[htb]
\begin{minipage}[t]{86mm}
\begin{center}
%\framebox[79mm]{\rule[-26mm]{0mm}{52mm}}
\epsfig{file=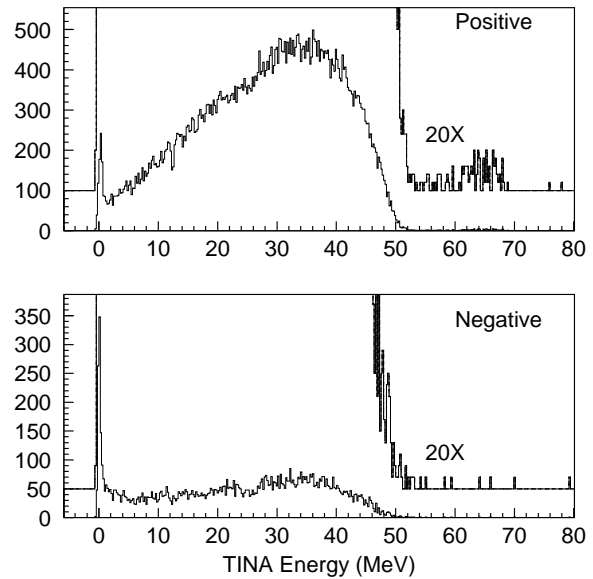,width=8cm,clip=}
\caption{Energy spectra in TINA
 for $\pi^+$ (top) and  $\pi^-$ (bottom)
stopping in H$_2$O.
Events in the energy region of 60--70 MeV are mainly 
from $\pi \rightarrow e \nu$ decays.}
%\label{fig:largenenough}
\end{center}
\end{minipage}
\end{figure}

\begin{table}
\begin{center}
\begin{tabular}{|l|r|l|r|l|l|}
\hline
 & \multicolumn{2}{c|}{$\pi^-$} &  \multicolumn{2}{c|}{$\pi^+$} & Free decay \\
Material & \multicolumn{1}{l}{$\pi \rightarrow e \nu$} 
& \hspace*{0.15cm} $\pi$ stops
 & \multicolumn{1}{l}{$\pi \rightarrow e \nu$} &
\hspace*{0.15cm} $\pi$ stops & 90 \% C.L.\\
\hline
H$_2$O & 3 & $1.29 \times 10^9$ & 69 & $5.23 \times 10^7$ & $4.6 \times 10^{-3}$ \\
Be & 1 & $8.32 \times 10^8$ & 94 & $5.40 \times 10^7$ & $3.1 \times 10^{-3}$ \\
Al & 1 & $3.34 \times 10^8$ & 114 & $6.38 \times 10^7$ & $7.6 \times 10^{-3}$ \\
Vinegar & 0 & $4.51 \times 10^8$ & 69 & $5.23 \times 10^7$ & $4.5 \times 10^{-3}$ \\
\hline
\end{tabular}
\end{center}
\caption{Summary of  $\pi \rightarrow e \nu$ analysis. The 90 \% C.L.
upper limits of the free decay fractions are
 shown in the last column for the case 
when $\tau_{obs} = \tau_{\pi}$.}
\end{table}

\subsection{C. Analysis of capture products}

In the search for free $\pi^-$ decay
via detection of delayed protons from $\pi^-$ capture,
 good timing is essential to be sensitive
to fast decay components. In order to minimize the time smearing, 
the energy loss as well as the TOF
cuts  in the selection of pions (typically $\pm 3 \sigma$ away from the peak)
were tighter than those used in the electron
analysis. 
Selection of protons was made based on the energy loss in all the three T counters;
the threshold for each counter was set around four times of that of minimum ionizing
particles. 
The resulting cut-off energy
 for accepted protons was around 45 MeV in the incident energy.
The contributions from  electron and muon contaminations in the beam
were negligible.
\begin{figure}[htb]
\begin{minipage}[t]{86mm}
%\begin{center}
%\framebox[150mm]{\rule[-26mm]{0mm}{52mm}}
\epsfig{file=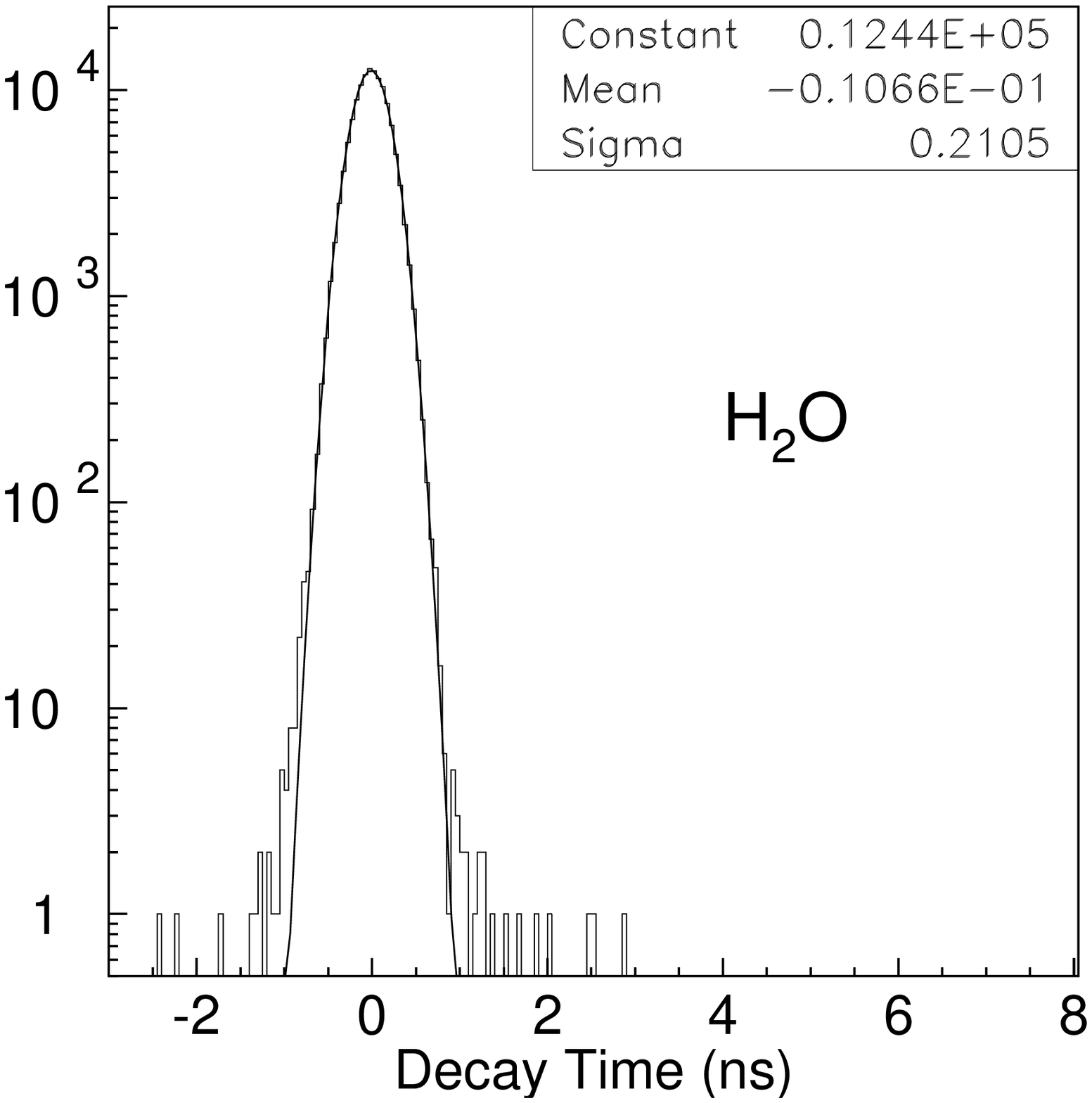,width=4.2cm,clip=}
\epsfig{file=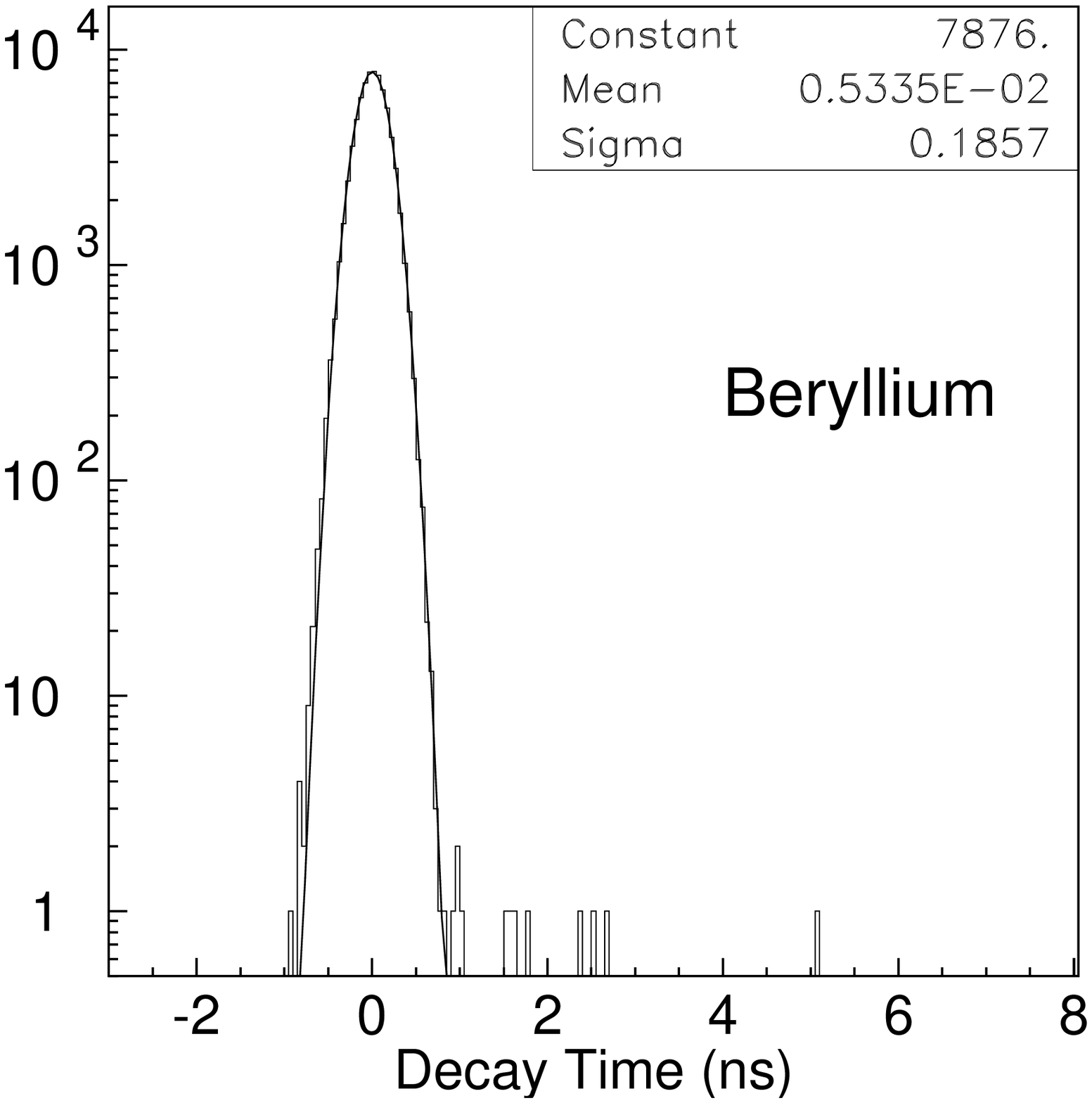,width=4.2cm,clip=}

\vspace*{0.2cm}

\epsfig{file=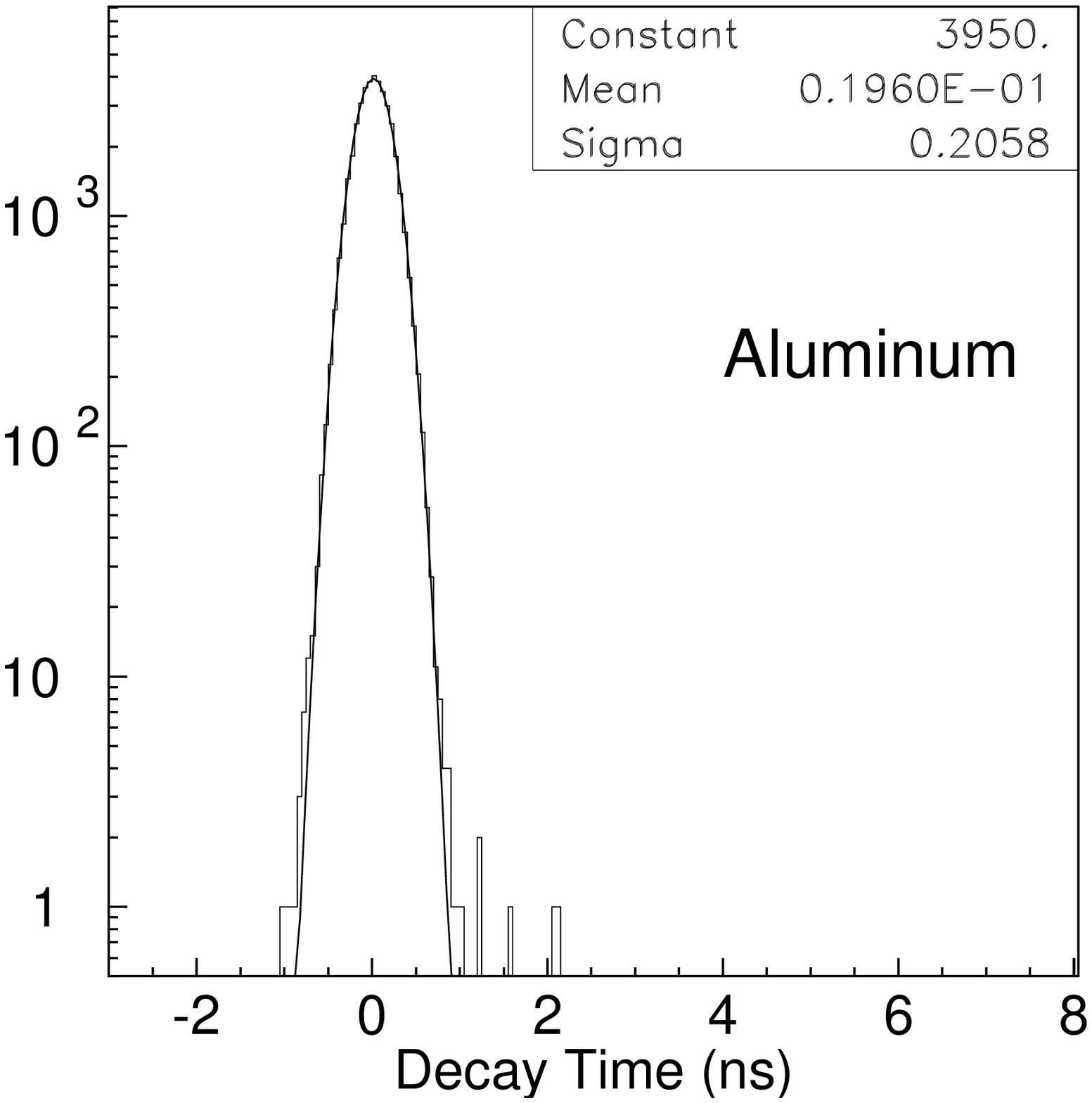,width=4.2cm,clip=}
\epsfig{file=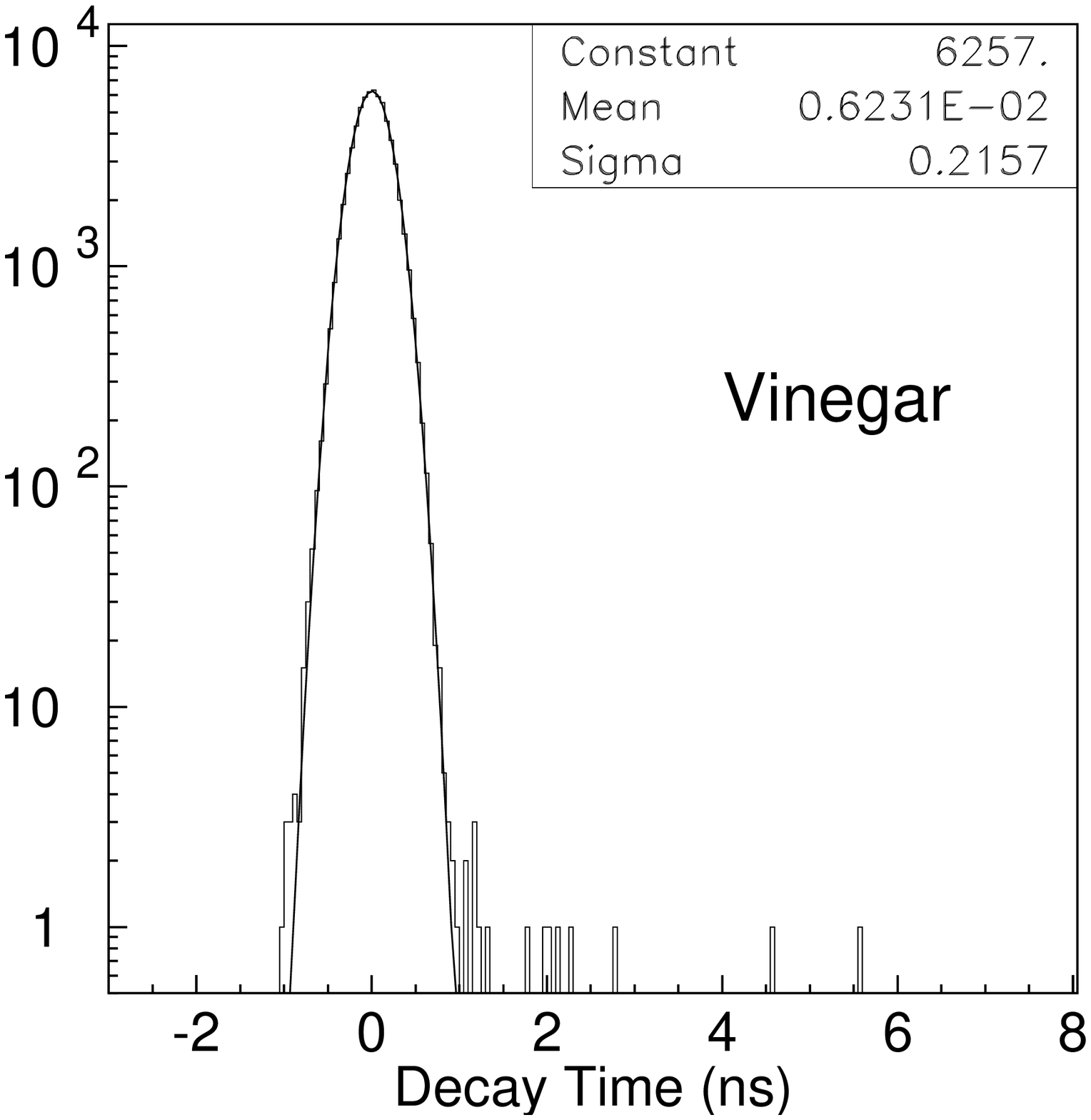,width=4.2cm,clip=}
\caption{Time spectra of protons for H$_2$O, Be, Al and vinegar
(histograms). The solid curves show Gaussian fits.}
%\label{fig:largenenough}
%\end{center}
\end{minipage}
\end{figure}

Without corrections the time resolution
of the prompt peak was $\sigma \sim$1 ns.
There were many effects that caused timing shift and spread; 
1) the beam momentum distribution
that affected the energy loss in the beam counters, the horizontal stopping
position, and the TOF, 2) the pulse height dependences
(time walk) in the beam counters and telescope counters, which
were correlated with the energy loss and the proton emission angle,
3) the path length variation, $e.g.$ due to the pion stopping position and the emission 
angle of protons, and 4) the propagation
time of the light in the scintillator especially in telescope
counters, T1, T2, and T3. Many effects were correlated with one another.
Correction factors were obtained by minimizing the dependence
of the $t_0$ on each measured parameter, such as pulse heights, timing
in the beam counters and telescope counters,
and horizontal pion stopping positions. A linear dependence was
used to obtain corrections. New correction factors were calculated 
at the end of each iteration and the revised correction factors
were applied in the next iteration (only half of the amount was corrected
each time). The process converged after about ten iterations.
The corrections improved the timing resolution
to $\sigma \sim 200$ ps.
These corrections do not affect the amount of the real decay component 
since the free decay component fraction is independent of the parameters 
used in the optimization.\\

\begin{figure}[htb]
\begin{minipage}[t]{86mm}
\begin{center}
\epsfig{file=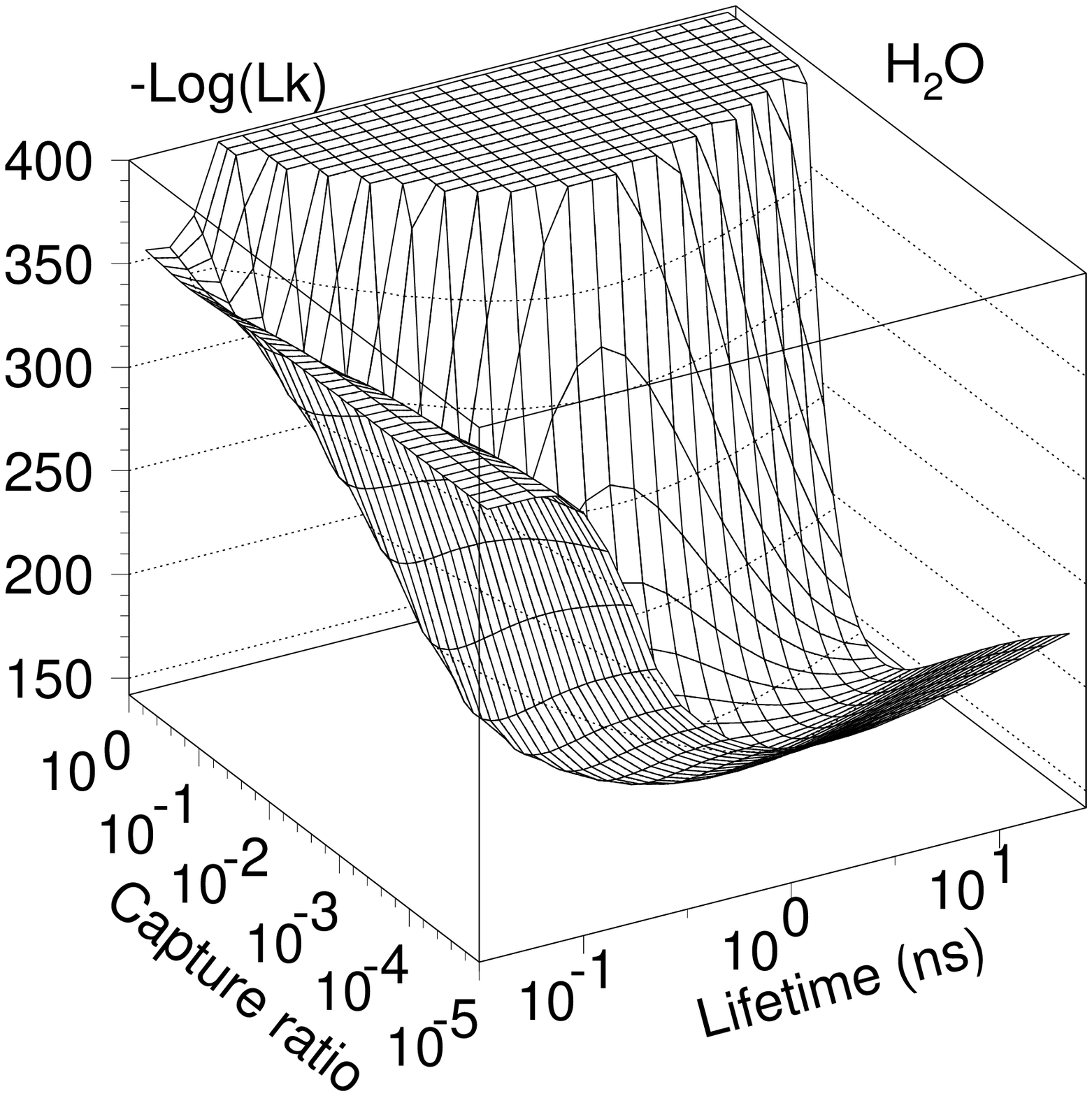,width=4.2cm,clip=}
\epsfig{file=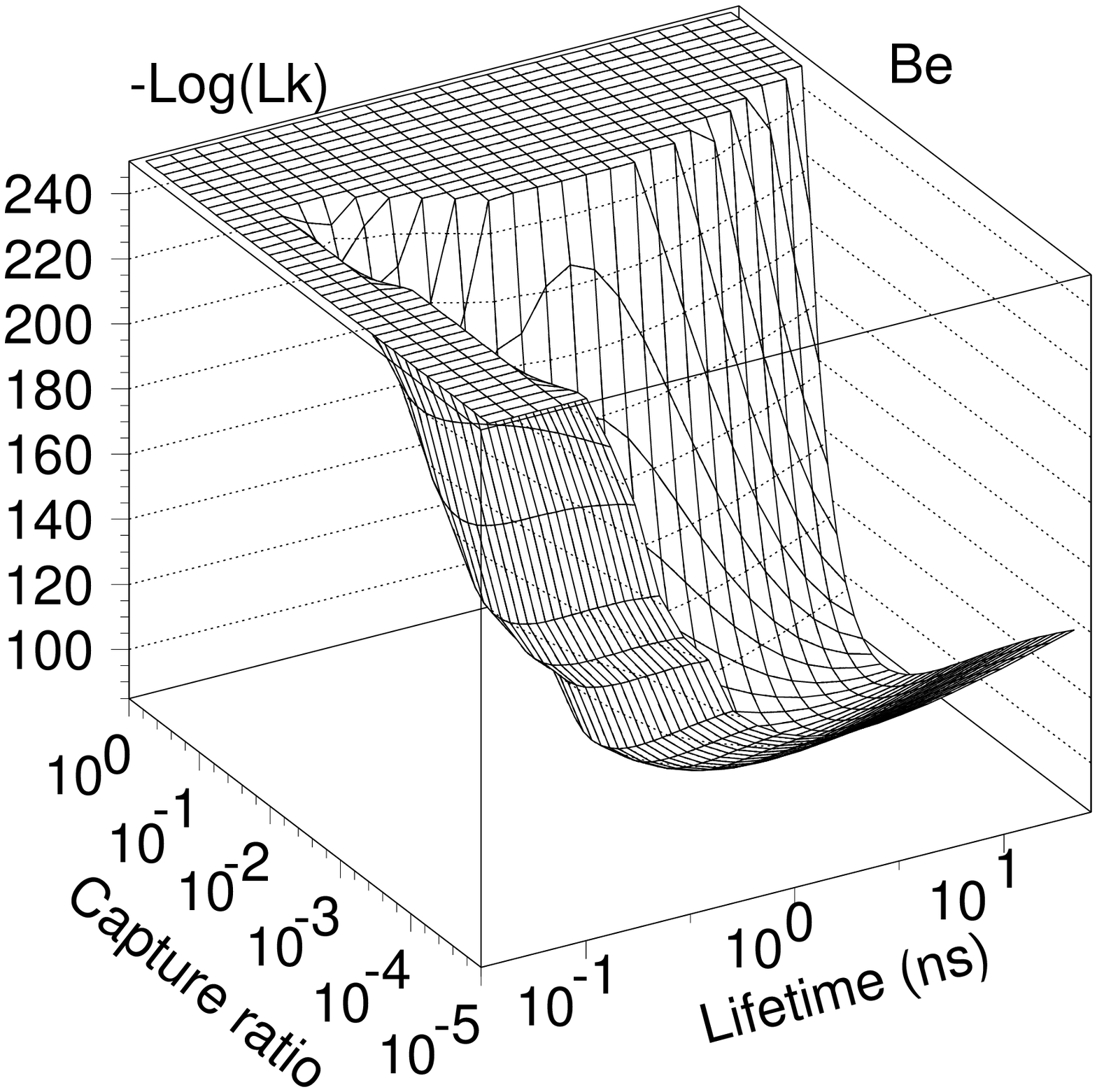,width=4.2cm,clip=}
\caption{-Log(Lk) for H$_2$O and Beryllium. Large  $- \log Lk$ values
were truncated for display.}
\end{center}
\end{minipage}
\end{figure}

 Figure 7 shows time spectra for H$_2$O,
Beryllium, Aluminum and vinegar. 
The time spectra of protons for all targets show essentially  a single component,
a Gaussian peak at $t_0$, as indicated by the solid line 
for a single Gaussian fit with three parameters; the amplitude, offset and
width, reflecting the resolution.
In the data from early stage of the experiment,
there was an obvious delayed component, a bump
at $t_p$=1--2 ns
that was enhanced when the events stopping in the upper half of the target
were selected.
The appearance of the delayed component coincided with the runs
with the water target filled only up to 85 \% of the height.
These suggested some slow pions
stopped outside the target (probably downstream of the target)
 with a longer flight path for the resulting proton. MC calculations
indicated that, when the target was 
lowered by the half height of the beam,
 about 0.3 \% pions stopped on the surface
of the material around the telescope counters
 and had a similar time distribution. 
 Since the extra
counts outside the Gaussian distribution could not
positively be identified as background,
we  treated them as a potential signal and only upper limits
are quoted.
\\

\begin{figure}[htb]
\begin{minipage}[t]{86mm}
\begin{center}
\epsfig{file=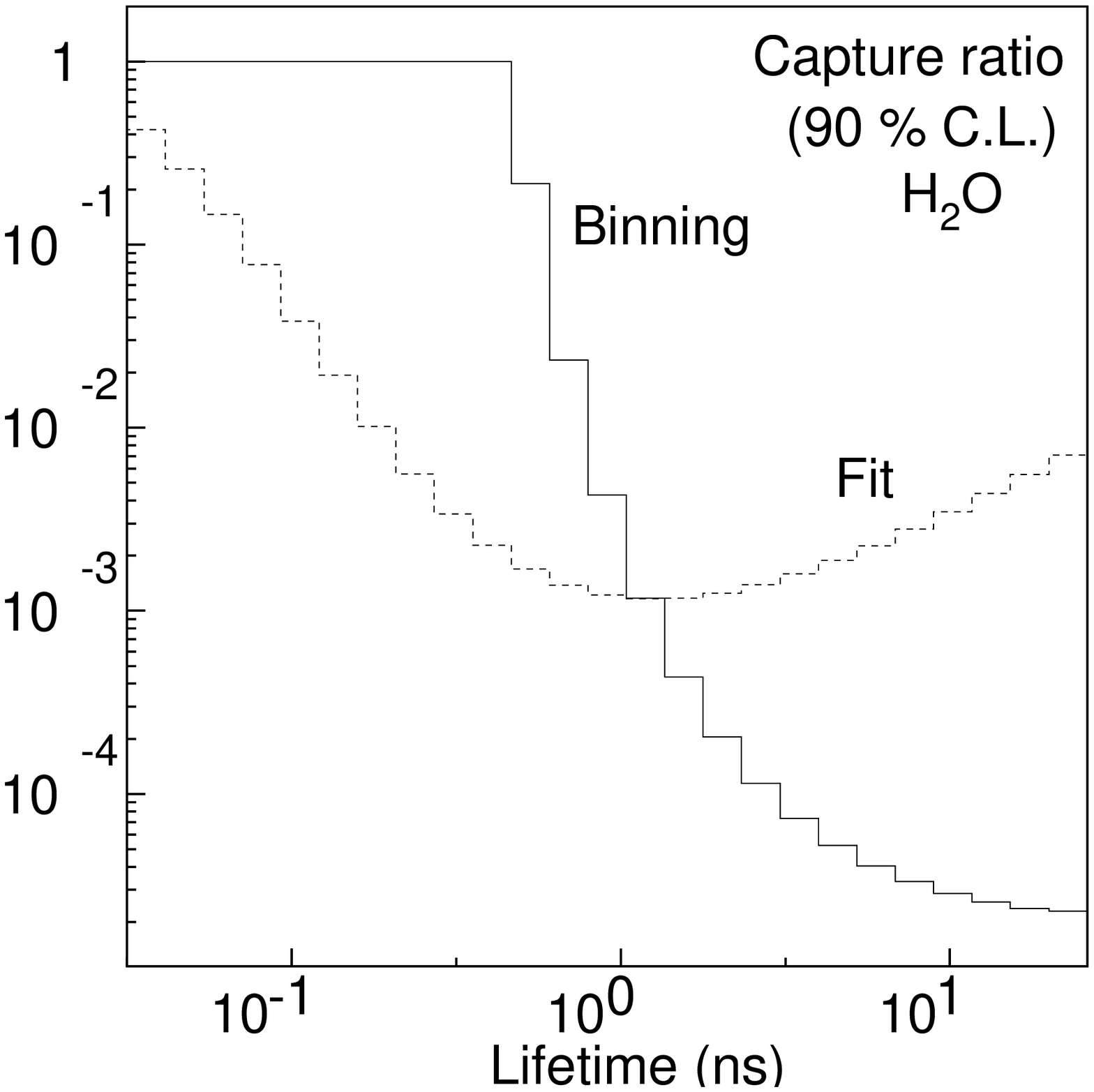,width=4.2cm,clip=}
\epsfig{file=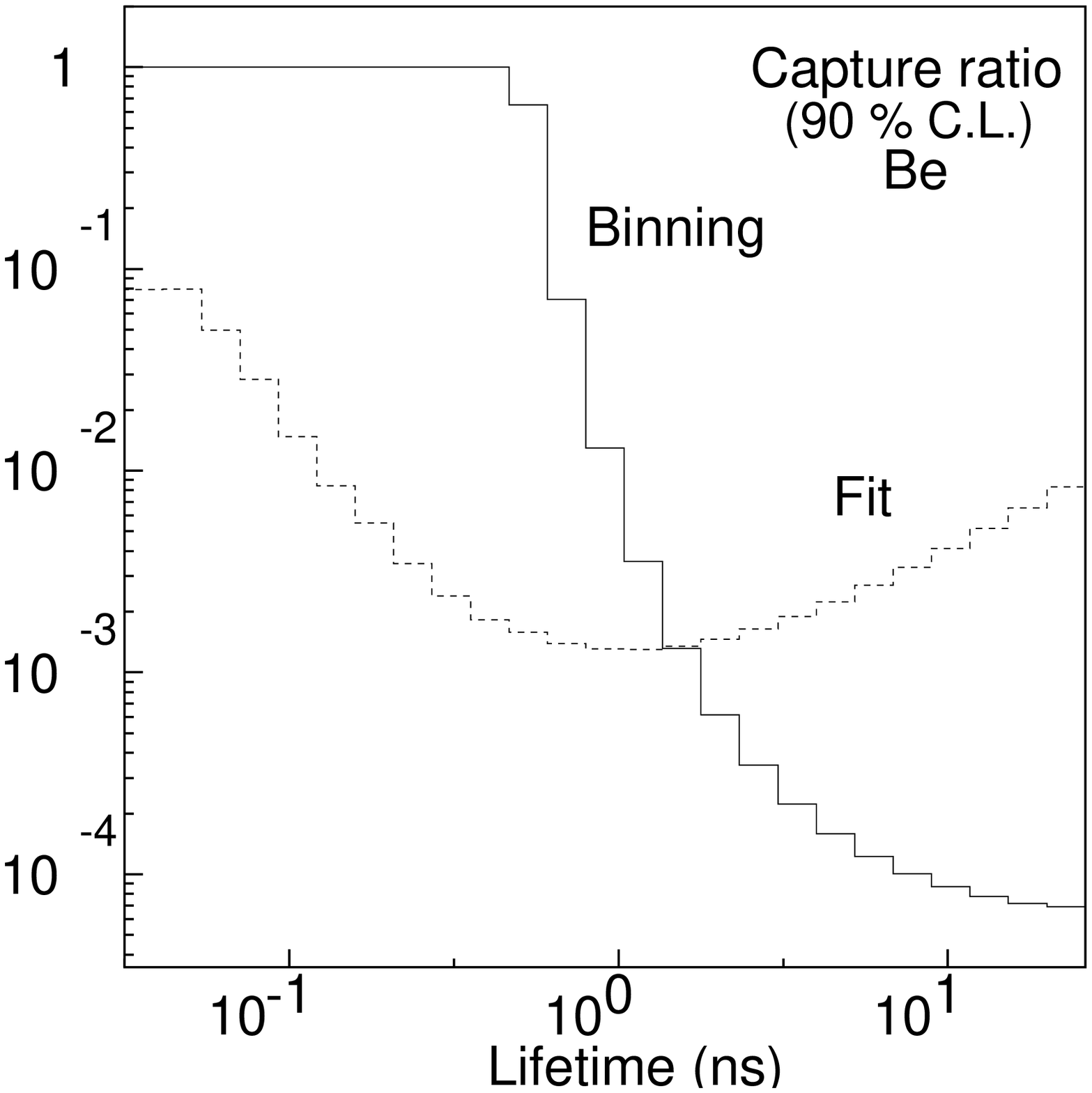,width=4.2cm,clip=}
\caption{90 \% C.L. upper limits of delayed capture ratios $f_{capt}$ 
for H$_2$O and Beryllium
as a function of assumed lifetime.
The true 90 \% C.L. limits around the 1-ns region may lie along a curve
that smoothly connects the binning and fit results.}
\end{center}
\end{minipage}
\end{figure}

For the extraction of a possible delayed component by the time spectrum fit,
the function used in the fit was a Gaussian
plus an exponential function including the
resolution effect (the same resolution was used). 
Assuming only one decay constant for the delayed component,
a likelihood function $Lk$
\begin{equation}
 Lk = \prod_{i=1}^{n}\frac{N_i^{n_i} e^{-N_i}}{n_i !} / 
\prod_{i=1}^{n}\frac{n_i^{n_i} e^{-n_i}}{n_i!},
\end{equation}
where $n_i$ and $N_i$ are the data and fitted value for the $i$-th bin, respectively,
 was maximized with three variable parameters;
the total counts, the resolution, and the $t_0$, but the lifetime and the capture ratio
of the delayed component were fixed for each maximization process, and the
minimum in $- \log Lk$ was stored for each set of lifetime and 
delayed capture ratio.
Fig. 8 shows surface plots of the likelihood functions $- \log Lk$ for the 
delayed capture ratio and the lifetime.
The dotted histograms in Fig. 9 show 
the 90 \% C.L. upper limits of the delayed capture ratio $f_{capt}$ for each lifetime.
\\

In order to test the validity of using a single Gaussian distribution
for the prompt capture timing peak,
a possible asymmetric time distribution, 
due to $e.g.$ solid angle effects causing different weight for different TOF,
 was simulated by MC.
Capture products were generated as soon as they stopped; pion capture in flight
was simulated separately.
Proton and deuteron spectra were taken from Ref.\cite{proton} and interpolated in energy
and atomic number.
Other heavier particles were ignored because of higher stopping power
that prevented them to contribute to a trigger.
The MC events were calibrated in the same way 
as the experimental data to reflect the effects
of the corrections in timing spectrum, and analyzed with the same
program with the same cuts.
After smearing with a resolution of 200 ps (the bare resolution of
MC was about 60 ps after applying corrections), the spectrum
 was fitted to a Gaussian curve.
 The fit agreed at the $10^{-4}$--$10^{-5}$ level,
and the asymmetry was not significant.

The effect of capture-in-flight was simulated 
with the total cross section \cite{rms},
$\sigma_{total} = \pi r_0^2 \cdot (1 + \frac{E_C}{T_{\pi}})$, where
$r_0 = 1.25 \cdot A^{1/3}$ $(fm)$ is the nuclear radius, 
and $E_C = 1.44 \cdot Z / r_0$ (MeV) is the Coulomb potential. $A$ and $Z$
are the mass and atomic numbers of the target material (Hydrogen was ignored
for scintillator and water), respectively. 
The nuclear absorption cross section was assumed to be 1/3
of the total cross section, and the capture products were generated
by ignoring the kinetic energy of the pion.
Inclusion of the capture-in-flight process increased the number of events
in the early time region ($t_p = -0.5$~ns) by 0.1 \% of the total counts,
which was consistent with the excess counts in the same time region of the data.
However, the capture ratio of the delayed component was unaffected by this. 
\\

In the long lifetime region, the fit results were affected by the small sample
bias of the Maximum Likelihood Method. A simple time bin method was employed
for this region; events in the time bin of $t_p = 5 - 75$ ns were considered to be
possible candidates, and the sum in the time bin was normalized
to the total number of events in the proton spectrum. A  Poisson distribution was
assumed to obtain the 90 \% C.L. upper limit. Table III summarizes the results
for H$_2$O, Be,  Al and vinegar. The solid lines in Fig.9 show the results
from the binning method.
\\

\begin{table}
\begin{center}
\begin{tabular}{|l|c|r|r|}
\hline
Material & Candidates & Total events & 90 \% C.L. ratios \\
\hline
H$_2$O & 0 & $1.31 \times 10^5$ &  $ 2.3 \times 10^{-5}$ \\
Be & 1 & $7.35 \times 10^4$ &  $ 6.9 \times 10^{-5}$ \\
Al & 0 & $4.08 \times 10^4$ & $7.3 \times 10^{-5}$ \\
Vinegar & 1 & $6.77 \times 10^4$ & $7.5 \times 10^{-5}$ \\
\hline
\end{tabular}
\end{center}
\caption{Summary of binning analysis of proton data.
The second column shows the numbers of candidate events
in the time bin 5--75 ns. The 90 \% C.L.
 upper limits for delayed capture
ratios in the last column are for 
$\tau_{obs} = \tau_{\pi}$ with the corresponding 
time-window correction of 0.769.}
\end{table}

\begin{figure}[htb]
\begin{minipage}[t]{86mm}
\begin{center}
%\framebox[150mm]{\rule[-26mm]{0mm}{52mm}}
\epsfig{file=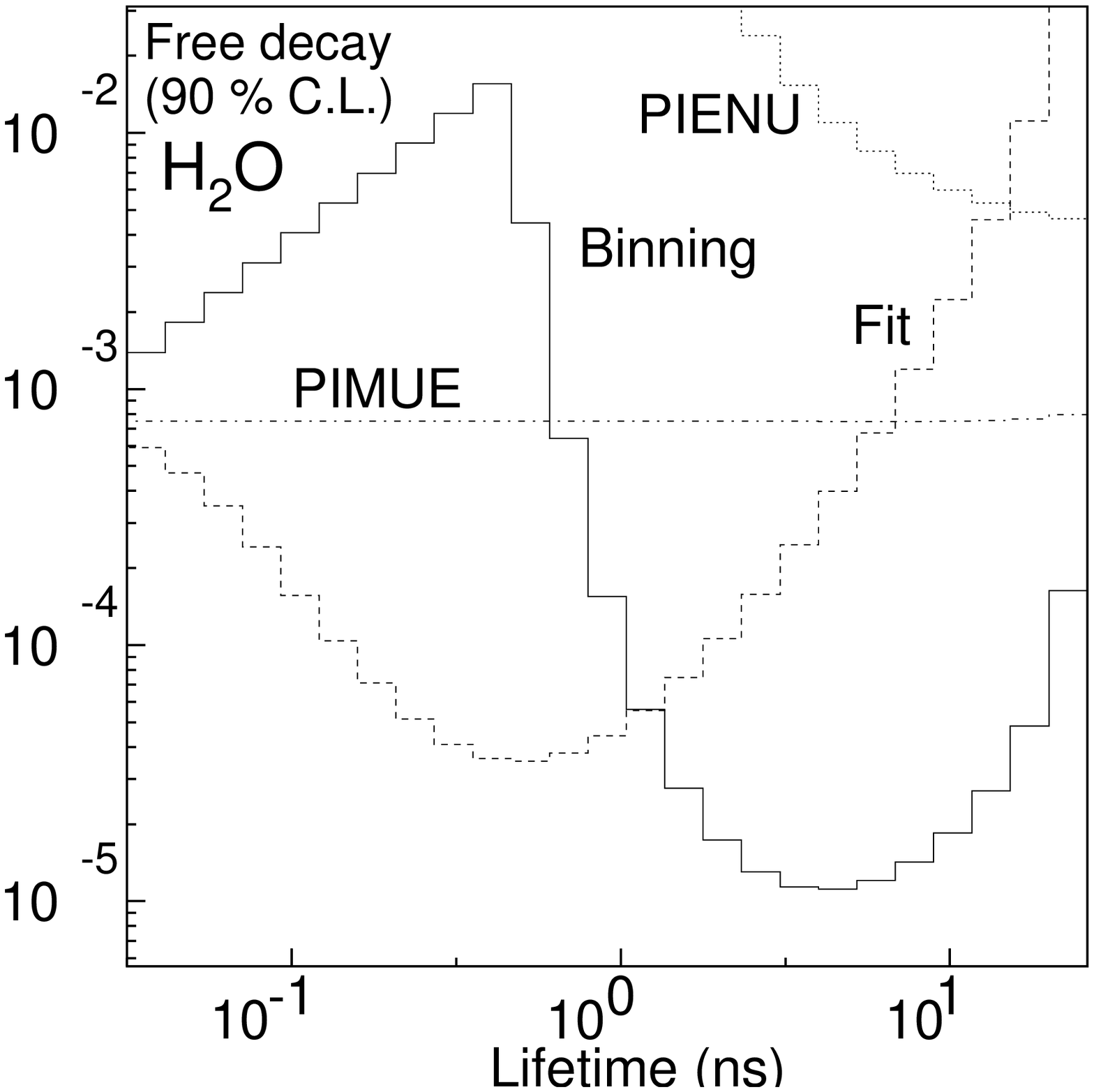,width=4.2cm,clip=}
\epsfig{file=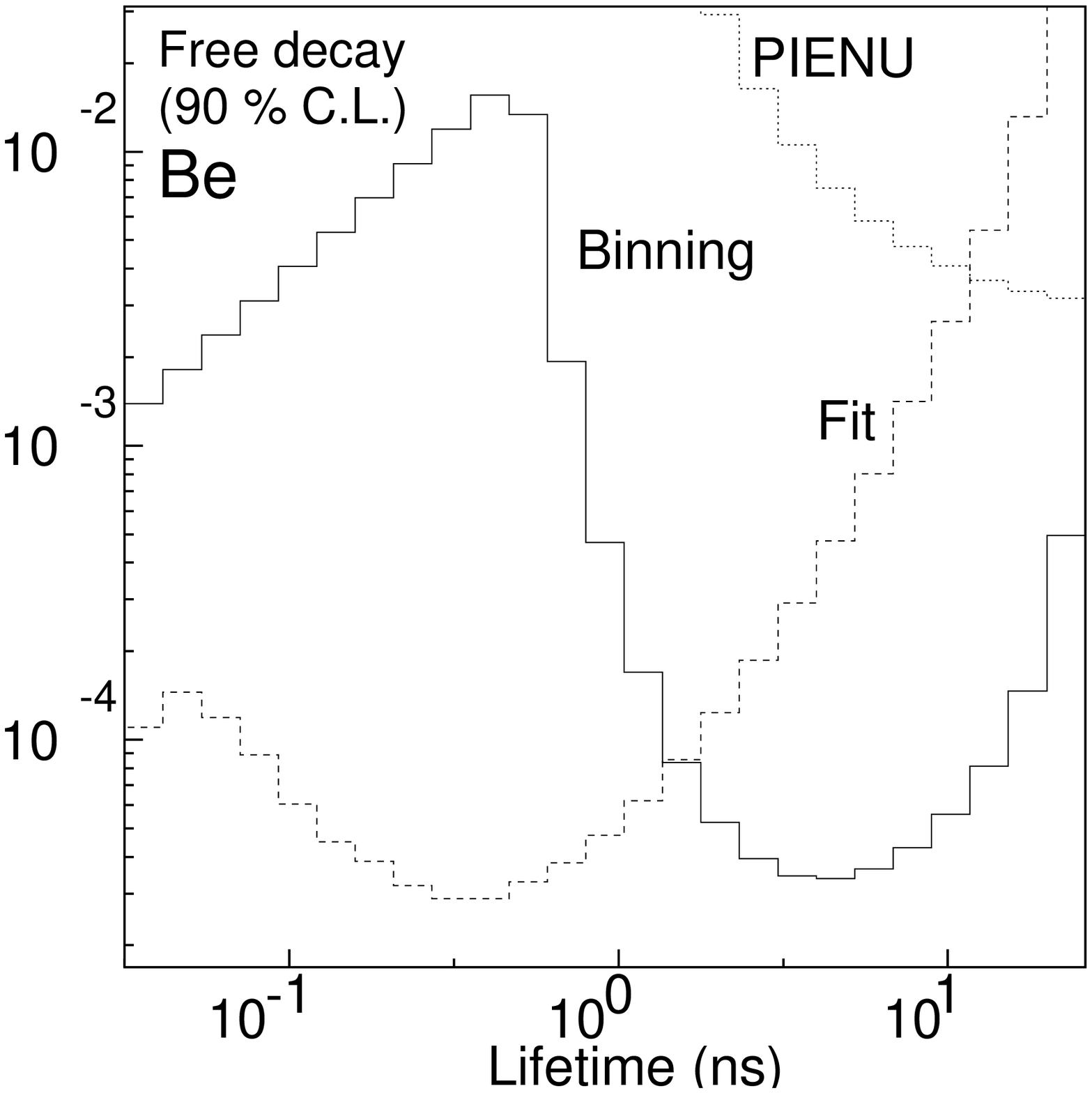,width=4.2cm,clip=}
\epsfig{file=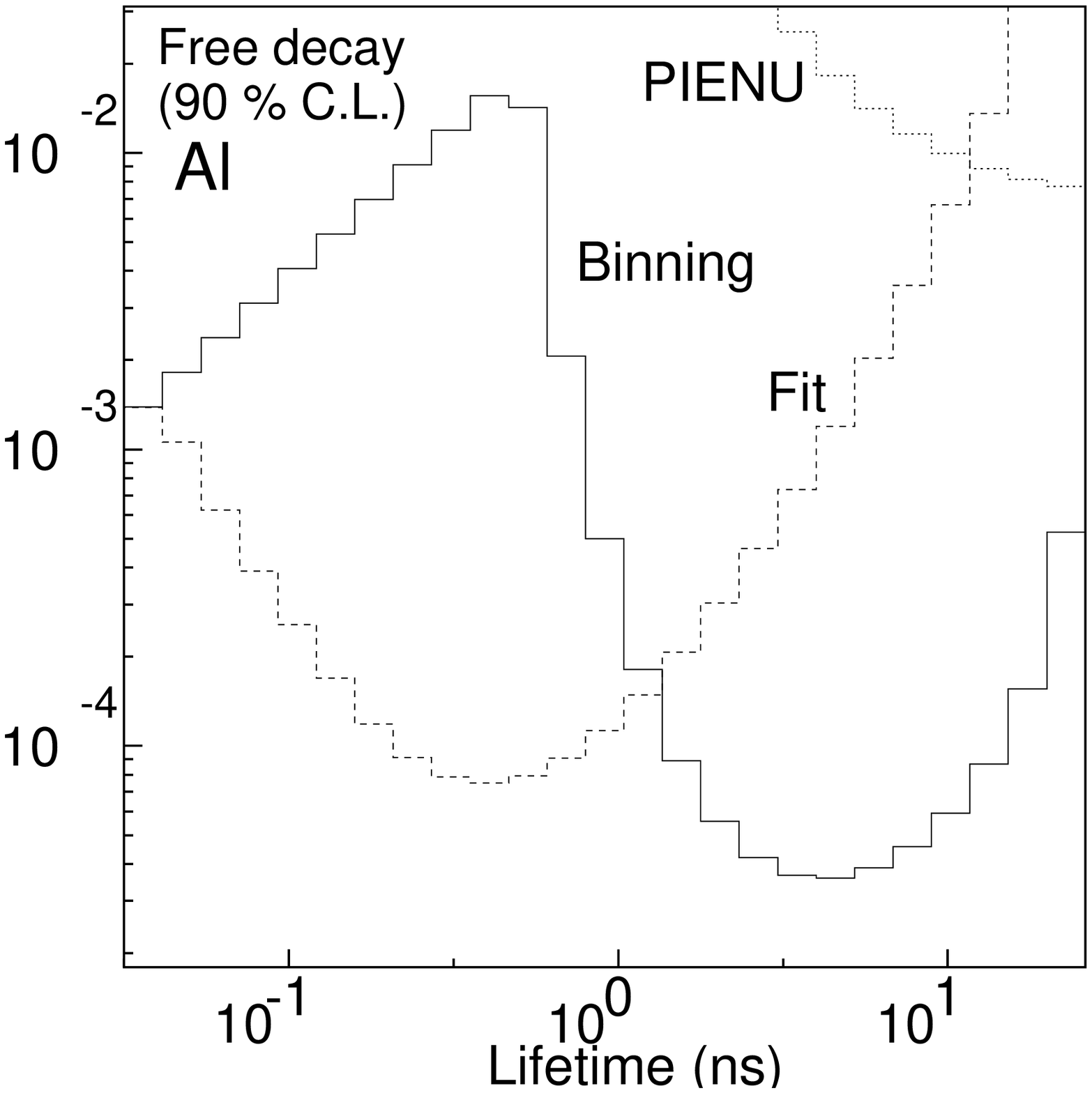,width=4.2cm,clip=}
\epsfig{file=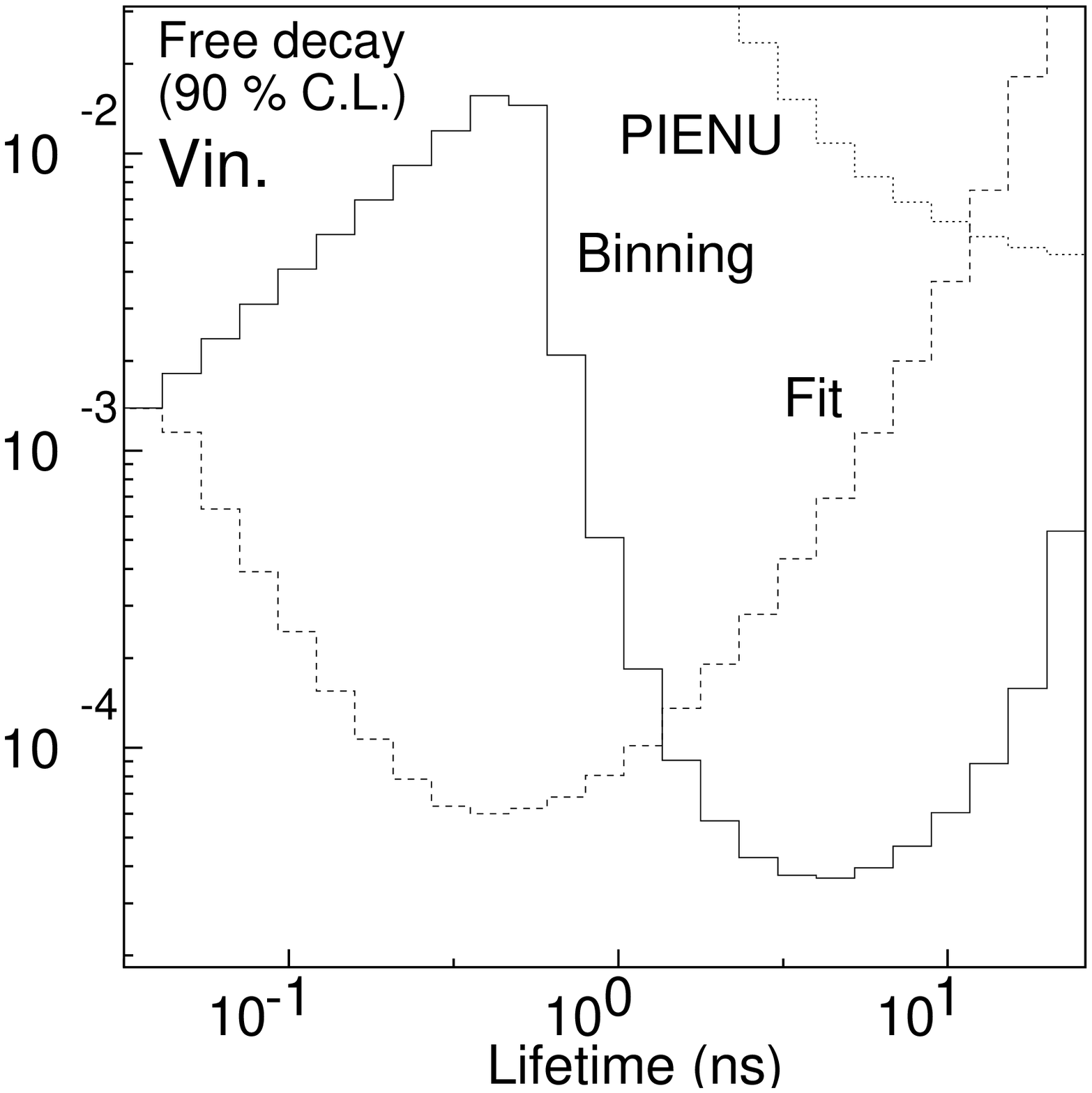,width=4.2cm,clip=}
\caption{90 \% C.L.upper limits of free decay fractions for H$_2$O, Be, Al, and vinegar.
The dashed histograms are the results by fit, the solid histograms
by binning for the delayed capture method,
 the dotted histograms by $\pi \rightarrow e \nu$ and
the dash-dotted line by $\pi \rightarrow \mu \rightarrow e$.}
%\label{fig:largenenough}
\end{center}
\end{minipage}
\end{figure}
%\hspace{\fill}

The observed delayed-capture ratio was converted to the fraction of
free $\pi^-$ decay using Eq.(2).
The summary of the proton analysis is shown in Fig. 10,
where the dashed lines indicate the 90 \% C.L. limits for the free decay
component obtained by the fits, and the solid lines by the binning method.
The capture limits go to unity at the pion lifetime (the next bin outside
the spectrum).
In the shorter lifetime region outside the plot,
the upper limit is along the extension of the binning result, which corresponds
to $\tau_{obs} / \tau_{\pi}$.
The results from the $\pi \rightarrow \mu \rightarrow e$ (dash-dotted line)
 and $\pi \rightarrow e \nu$ (dotted line)
analyses are also shown in the figure.
Without knowing the lifetime of the delayed component,
conservative limits obtained near the pion lifetime 
(the largest upper limits) for all cases are:
$8.2 \times 10^{-4}$ for H$_2$O by the $\pi \rightarrow \mu \rightarrow e$ analysis
at $\tau_{obs} = \tau_{\pi}$, and
$3.2 \times 10^{-3}$ for Beryllium,
$7.7 \times 10^{-3}$ for Aluminum and
$4.5 \times 10^{-3}$ for vinegar by $\pi \rightarrow e \nu$
at the intersection of the binning and $\pi \rightarrow e \nu$ limits.
\\

\section{V. Conclusion}

No free $\pi^-$-decay component was observed in water with the ``90 \% C.L.''
upper limit for the free decay fraction
of $8.2 \times 10^{-4}$  for the worst case ($\tau_{obs} = \tau_{\pi}$).
For Beryllium and Aluminum assuming a single delayed component, 
the limits were $3.2 \times 10^{-3}$
and $7.7 \times 10^{-3}$ coming from the measurements of
$\pi^- \rightarrow e^- \overline{\nu}$ decays, respectively.
The result of the present experiment indicates that the materials
  studied here will not cause $\overline{\nu}_e$ contamination in the
  beam dump neutrino experiments.
It seems  unlikely that the LSND result\cite{lsnd},
which would require the relative $\overline{\nu}_e$ yield of $> 10^{-2}$\cite{nue},
 can be explained by the mechanism
of unusual formation or de-excitation processes of pionic atoms in water.\\

\begin{acknowledgements}

The authors would like to express their gratitude to P. Amaudruz for
his help in setting up the data acquisition system.
This work was in part supported by the National Council of Canada,
the National Science Foundation, and the Devision of Nuclear Physics, Department
of Energy.
\end{acknowledgements}


\begin{thebibliography}{9}
\bibitem{early} The corresponding early time region $t_e = 5 - 38$~ns
suffers from the tail of prompt events. It is also less sensitive to slow 
decay components.
\bibitem{muatom} For the early $\mu$-decay time window between $t_1$ and $t_2$,
  the product of the muon-decay fraction $\tau_A / \tau_{\mu}$ and the
  time acceptance $e^{-t_1 / \tau_A} - e^{-t_2 / \tau_A}$
  can be approximated to  $e^{-t_1 / \tau_{\mu}} - e^{-t_2 / \tau_{\mu}}$,
  where $\tau_A$ is the lifetime of the muonic atom.
  The advantage of this approach over the use
  of muonic atom is that the contributions from $\mu$-decays in different
  materials can be added with a single lifetime.
\bibitem{nue} The signal to $\overline{\nu}_e$-originated background ratio
was $\sim$15 with the expected $\overline{\nu}_e /\overline{\nu}_{\mu}$ 
flux ratio of $7.8 \times 10^{-4}$.
Considering  the $\pi^+/\pi^-$ production ratio and
the pion stopping ratio in the water target, 
the required free decay fraction to explain
the LSND result could be as high as 0.1--1.
\end{thebibliography}
\end{document}